
\input harvmac

\def \s{\sigma}
\def \b{\beta}
\def \a{\alpha}
\def \g{\gamma}
\def \d{\delta}
\def \e{\epsilon}
\def \l{\lambda}

\def \ph{\phi}

\def \Ph{\Phi}
\def \p{\pi}

\def \D{\Delta}
\def \G{\Gamma}

\def \m{\mu}
\def \n{\nu}

\def \O{\Omega}
\def \D{\Delta}
\def \r{\rho}
\def \k{\kappa}

\def\apny{Ann.\ Phys.\ (New York)\ }
\def\cmp{Comm.\ Math.\ Phys.\ }

\def\ijmpa{{Int.\ J.\ Mod.\ Phys.\ }{\bf A}}

\def\mpla{{Mod.\ Phys.\ Lett.\ }{\bf A}}
\def\nc{Nuovo Cimento\ }
\def\npb{{Nucl.\ Phys.\ }{\bf B}}

\def\plb{{Phys.\ Lett.\ }{\bf B}}

\def\pr{Phys.\ Rev.\ }
\def\prd{{Phys.\ Rev.\ }{\bf D}}
\def\prl{Phys.\ Rev.\ Lett.\ }

\def\zpc{Z.\ Phys.\ {\bf C}}
\def\lmp{Lett.\ Math.\ Phys.\ }
\def\ua{\underline{\a}}
\def\ud{\underline{\d}}
\def\uH{\underline{H}}
\def\ur{\underline{\r}}
\def \cH{{\cal H}}
\def \pp{\prime\prime}
\def \pri{\prime}
\def \tD{\tilde\D}
\def \tP{\tilde\Ph^+}
\def \trho {\tilde \r}
\def \tPm{\tilde\Ph^-}
\def \ba{\bar\a}
\def \ta{\tilde\a}
\def \tb{\tilde \b}
\def \del {\partial}
\def \bdel{\bar \del}
\def \bz {\bar z}
\def \cG {{\cal G}}
\def \tG {\tilde  G}
\def \tcG {\tilde{\cal G}}
\def \ti {\tilde \imath}
\def \tj {\tilde \jmath}
\def \bi {\bar \imath}

\def \tra {\tilde a}
\def \trb {\tilde b}
\def \bJ{\bar J}
\def \bT{\bar T}

\def \cJ{{\cal J}}
\def \ad{{\rm ad}}
\def \cC {{\cal C}}

\def \bs{\setminus}

\def \tg{\tilde g}
\def \bg{\bar \g}
\def \tB{\tilde B}
\Title{LTH 315}{Quantum Non-abelian Toda Field Theories}
\centerline{I. Jack, D. R. T. Jones and J. Panvel}
\bigskip
\centerline{\it DAMTP, University of Liverpool, Liverpool L69 3BX, U.K.}
\vskip .3in
We derive an explicit, exactly conformally invariant form for the action for
the non-abelian Toda field theory. We demonstrate that the conformal
invariance conditions, expressed in terms of the $\beta$-functions of the
theory, are satisfied to all orders, and we use our results to obtain a
value for the central charge agreeing with previous calculations.

\Date{August 1993}

\newsec{Introduction}
Toda field theories provide examples of conformal field theories with a rich
and interesting structure, and consequently have been investigated from
various points of view over a considerable period
\ref\toda {M. Toda, Phys. Rep. {\bf C}18 (1975) 1\semi
A.N. Leznov and M.V. Saveliev, Lett. Math. Phys.  3 (1979)
      489; Commun. Math. Phys. 74 (1980) 111\semi
P. Mansfield, Nucl. Phys. {\bf B}208 (1982) 277\semi
D. Olive and N. Turok, Nucl. Phys. {\bf B}220 (1983) 491\semi
J.-L. Gervais and A. Neveu, Nucl. Phys. {\bf B}224 (1983) 329\semi
E. Braaten, T. Curtright, G. Ghandour and C. Thorn, Phys. Lett.
{\bf B}125 (1983) 301\semi
O. Babelon, Phys. Lett. {\bf B}215 (1988) 523\semi
O. Babelon, F. Toppan, and L. Bonora, Commun. Math. Phys. 140 (1991) 93\semi
A. Bilal and J-L Gervais, Phys. Lett. {\bf B}206 (1988) 412; Nucl.
Phys. {\bf B}314 (1989) 646; {\bf B}318 (1989) 579\semi
P. Mansfield and B. Spence, Nucl. Phys. {\bf B}362 (1991) 294.}
\ref\pma {P. Mansfield, Nucl. Phys. {\bf B}222 (1983) 419.}
\ref\bill{G. Papadopoulos and B. Spence, ``The Space of Solutions of Toda Field
Theory'', preprint UM-P-93/38, KCL-93-7, hep-th/9306088.}.
The standard Toda field
theories are each associated with a Lie algebra; to be more precise, with the
canonical grading of a Lie algebra, in which the Cartan subalgebra furnishes
the zero-grade part of the algebra, and the positive and negative grade
components are generated by the step operators corresponding to
positive and negative roots respectively.
More generally, one can associate a generalised Toda field theory with an
arbitrary integral
grading of a Lie algebra, one for which the zero-grade component
contains some of the step operators in addition to the Cartan subalgebra
generators. Such theories were first considered in Ref. \ref\lsI{A. N.
Leznov and M. V. Saveliev, \lmp6 (1982) 505; \cmp89 (1983) 59.}, where
however the authors were principally interested in integral gradings
corresponding to integral embeddings of $sl(2)$ into the Lie algebra. More
general integral gradings have been considered in Refs. \ref\OW
{L. O'Raifeartaigh and A. Wipf, \plb251 (1990) 361.}\ref\ORTW{
L. O'Raifeartaigh, P. Ruelle, I. Tsutsui and A. Wipf, \cmp143 (1992)
333.}\ref\FORTW{L. Feh\'er,
L. O'Raifeartaigh, P. Ruelle, I. Tsutsui and A. Wipf,
\apny213 (1992) 1.}, and it is these which we shall describe. (Half-integral
gradings are discussed in Ref. \ref\FORTWi
{L. Feh\'er, L. O'Raifeartaigh, P. Ruelle, I. Tsutsui and A. Wipf,
Phys. Rep. 222 (1992) 1.}.)
Such generalised Toda theories are termed ``non-abelian''
since the zero-grade component is now non-abelian. The Lagrangian for the
standard, or abelian
Toda field theory is characterised by a set of fields with standard kinetic
terms and simple exponential interaction terms; the non-abelian Toda theories,
on the other hand, have a non-linear $\s$-model type of kinetic term
( in fact, the action for a Wess-Zumino-Witten model
\ref\ew{E. Witten, \cmp92 (1984) 483.}) together with more
complex interactions featuring polynomials multiplying exponentials.
Non-abelian Toda theories were first introduced some time ago, but have
recently been receiving rather more attention\ref\sav{M. V. Saveliev,
\mpla5 (1990) 2223.}
\ref\gs{J.-L. Gervais and M. V.
Saveliev, \plb286 (1992) 271.}
\ref\dgs{F. Delduc, J.-L. Gervais and M. V. Saveliev, \plb292 (1992) 295.}
\bill. In particular it has been
shown how they can be derived by Hamiltonian reduction of the
Wess-Zumino-Witten (WZW) model\OW. (This reduction can also be implemented by
gauging the WZW model, as discussed in detail in the abelian context in Ref.
\ref
\BFOFWI{J. Balog, L. Feh\'er, L. O'Raifeartaigh, P. Forg\'acs and
A. Wipf, \apny203 (1990) 76.}.)
By virtue of this relation with the WZW model, the
non-abelian Toda theories should be exactly conformally invariant at the
quantum level, just as for the ordinary Toda theory, and this can be shown by
construction of the energy-momentum tensor for the non-abelian Toda theory
from that for the corresponding WZW model\ref\BFOFWII{J. Balog, L. Feh\'er,
P. Forg\'acs, L. O'Raifeartaigh and A. Wipf, \plb227 (1989) 214.}\OW.
Our goal in this paper is to show
explicitly how to construct an exactly conformally invariant action for the
non-abelian Toda theory. The classical action for the
non-abelian Toda theory, obtained from the WZW action by the reduction
process,
 is of the non-linear $\s$-model type\OW, and the
conditions for such an action to be conformally invariant have a well-known
formulation
\ref\betab{A. A. Tseytlin, \plb178 (1986) 34; \npb294 (1987) 383\semi
G. M. Shore, \npb286 (1987) 349.}
\ref\ho{H. Osborn, \npb294 (1987) 595.}
in terms of the renormalisation-group $\b$-functions for the
theory
\ref\betaa{D. Friedan, \prl51 (1980) 334; \apny163 (1985) 318.}.
We shall show how these conditions may be satisfied to all orders
by adjusting the couplings in the classical action and also adding to the
action a
dilaton field, coupling to the two-dimensional scalar curvature. We
corroborate our results by showing that we reproduce the known result for the
central charge\OW\ for the non-abelian Toda theory.

The essence of our method is to identify the conformal invariance
condition for the scalar potential in the non-abelian Toda action,
expressed in terms of the $\b$-function, with the Virasoro condition for the
scalar potential to be a primary field (or rather, to be precise, with the
zeroth order term in the Laurent expansion of this condition). The Virasoro
condition can be given explicitly in terms of the Casimir operator for the
Lie algebra associated with this non-abelian Toda theory; on the other hand,
the conformal invariance condition can be determined explicitly in terms of
the metric and dilaton fields for the $\s$ model, exploiting the fact that the
kinetic terms are those of a WZW model, up to multiplication by constants.
Thus we can identify these constants and the required dilaton field.
This procedure has elements in common with ideas used in
discussing quasi-exactly-soluble
quantum mechanical systems
\ref\tur{A. V. Turbiner, \cmp118 (1988) 467; M. A. Shifman
and A. V. Turbiner, \cmp126 (1989) 347; M. A. Shifman, \ijmpa4 (1989) 2897.}
and the energy-momentum tensor of conformal field
theories based on the generalised Sugawara construction
\ref\halp{M. B. Halpern and
E. B. Kiritsis,\mpla4 (1989) 1373;\mpla4 (1989) 1797(E).}.
Recently it has been used very successfully
to derive exact metric
and dilaton fields for string black holes\ref\dvv
{R. Dijkgraaf, E. Verlinde
and H. Verlinde, \npb371 (1992) 269.}
\ref\bars{I. Bars and K.
Sfetsos, \prd46 (1992) 4495, 4510; \plb301 (1993) 183.},
yielding results which, in the case
of Witten's original string black hole solution
\ref\sbh{E. Witten, \prd44 (1991) 314.}
, have been checked to fourth
order in perturbation theory
\ref\jjp{A. A. Tseytlin, \plb268 (1991) 175; I. Jack, D. R. T. Jones and
J. Panvel, \npb393 (1993) 95.}. The technique has already been used to
obtain the central charge and dilaton for the ordinary abelian Toda theory
\ref\jp{I. Jack and J. Panvel, ``WZW-Toda Reduction using the Casimir
Operator'', Liverpool preprint LTH304, hep-th 9302077 (\ijmpa,
to be published)},
and we shall refer back to this paper for a fuller explanation of some of
the results we shall use. In the abelian case it is possible to derive the
results more quickly by assuming the general form of the Toda action and then
simply solving the conformal invariance conditions, and this was in fact
done some time ago
\ref\arkI{A. A. Tseytlin, \plb241 (1990) 233.}; however, this would be
more difficult in the non-abelian case without the guidance provided by the
knowledge that the scalar potential satisfies the Virasoro condition.

The plan of the paper is as follows: in Section 2 we define non-abelian Toda
field theories, focussing on their derivation by reduction of a WZW model.
In Section 3 we discuss the conformal invariance conditions for a field theory
in terms of the renormalisation group $\b$-functions for a non-linear
$\s$-model,while in Section 4 we obtain a differential equation for the
potential
term in the non-abelian Toda theory from the Virasoro condition.
In Section 5 we compare the $\b$-function condition for the potential with
this differential equation, which enables us to read off the exact quantum
form of the action for the non-abelian Toda theory. As a check we derive the
central charge for the theory. In Section 6 we exemplify the above
considerations with reference to the particular case of a non-abelian Toda
theory based on the Lie algebra $B_2$. Finally in Appendix A we define our
notation and conventions for Lie algebras, and prove two theorems in group
theory which we need in the main text.

\newsec{Non-abelian Toda theories}
In this section we define non-abelian Toda theories and fix our
notation. Non-abelian Toda theories were first introduced\lsI by
generalising the Lax pair representation for a conventional Toda theory
(which uses the canonical grading for the associated Lie algebra) to
the case of a Lie algebra with arbitrary grading. However, we shall find it
more convenient here to obtain the non-abelian Toda theory by
Hamiltonian reduction\OW\FORTW (or equivalently gauging\BFOFWI)
of a Wess-Zumino-Witten
model. We start by discussing the notion of grading a Lie algebra
$\cG$.
We suppose that the Lie algebra $\cG$ has a set of simple roots
$\a\in\D$ and a
corresponding set of positive roots $\Ph^+$.
We define the dual simple roots (or fundamental weights) $\a^{\pri},\quad
\a\in\D$
to satisfy
\eqn\eIi{
2{\b^{\pri}.\a\over{\a.\a}}=\d_{\a\b} }
so that $\a^{\pri}$ is given explicitly by
\eqn\eIii{
\a^{\pri}=\sum_{\b\in\D}A^{\a\b}\b, }
where $A^{\a\b}$ is the inverse of the Cartan matrix $A_{\a\b}$.
 We can introduce an integral grading of the
Lie algebra as follows: suppose we select a subset of the simple roots
$\tD\subset\D$.
It will also be convenient to define
$\tP$ to be the set of positive roots which are sums of $\ta,\quad
\ta\in\tD$, and $\tPm$ to be the corresponding negative roots.
We now define
\eqn\eIiii{
\d=2\sum_{\ba\in\D\bs\tD}{\ba^{\pri}\over{\ba.\ba}}.  }
The corresponding element $\d.H=\d_iH_i$
of the Cartan subalgebra now acts as a
grading
operator; $\d$ has the properties
\eqn\eIiiia{\eqalign{
\d.\ta&=0,\qquad\ta\in\tD\cr
\d.\ba&=1,\qquad\ba\in\D\bs\tD\cr} }
from which follows
\eqn\eIiv{
\eqalign{
[\d.H,E_{\ta}]&=0,\qquad \ta\in\tD, \cr
[\d.H,E_{\ba}]&=E_{\ba},\qquad \ba\in\D\bs\tD. \cr}}
We may now decompose $\cG$ into $\cG_{\pm}$, $\cG_0$
where $X\in \cG$ is assigned to $\cG_+$, $\cG_-$ or $\cG_0$ according as
$[\d.H,X]=nX$ where the integer $n$ is positive, negative or zero
respectively. Hence $\cG_0$ is generated by $E_{\pm\ta},\quad\ta\in\tP$ and by
the Cartan subalgebra generators.
This is not the most general integral grading possible, since
with our definition the simple roots are always assigned a grade of one or
zero; this is necessary\FORTW\ to ensure the existence of the
Drinfeld-Sokolov gauge\ref\ds{V. Drinfeld and V. Sokolov, J. Sov. Math. 30
(1984)
1975.}\ and hence to guarantee a polynomial realisation of
the Kac-Moody algebra (see later).
  It is convenient now to select a basis for the
Cartan subalgebra which is in accord with the decomposition. We start with
the Cartan subalgebra generators $\ta.H,\quad\ta\in\tD$. These
span some $r^{\tcG}$-dimensional subspace of the Cartan subalgebra, where
$r^{\tcG}$ is the number of simple roots in $\tD$.
We select an orthonormal basis $H_i,\quad i=1,2,\ldots r^{\tcG}$ for this
subspace. We then extend this basis to an orthonormal basis
$H_i,\quad i=1,2,\ldots r^{\cG}$ (where $r^\cG$ is the dimension of $\cG$)
for the whole Cartan subalgebra.
We then have
\eqn\eIiva{
\ta_{\bi}=0, \qquad [H_{\bi},E_{\ta}]=\ta_{\bi}E_{\ta}=0
\quad(\bi=r^{\tcG}+1,\ldots r^{\cG}).}
Then we see from \eIiva\ that $\cG_0$ itself splits into a direct sum
of $\tcG$, the Lie algebra whose generators are $E_{\pm\ta},\quad\ta\in
\tP$ and $H_i\quad
i=1,2,\ldots r^{\tcG}$, together with $r^{\cG}-r^{\tcG}$ factors of $R$.
$\cG_+$ is generated by
$E_{\ba},\quad \ba\in\Ph^+\bs\tP$,
while $\cG_-$ is generated by
$E_{-\ba},\quad \ba\in\Ph^+\bs\tP$. This decomposition of $\cG$ is called a
grading of the Lie
algebra.
{}From now on we shall assume the index $\ti$ runs from $1$ to $r^{\tcG}$,
while the index $\bi$ will run from $r^{\tcG}+1$ to $r^{\cG}$. The index $i$
will be assumed to take values from $1$ to $r^{\cG}$.
We shall raise and lower indices, and perform
contractions, using the ordinary Kronecker delta.
As a special case of the grading of the Lie algebra
we may take $\tD=\emptyset$, in which case $\cG_0$
is simply the Cartan subalgebra and $\cG_{\pm}$ consist of the algebras
generated by $E_{\a}, \a\in\Ph^+$ and by $E_{-\a}, \a\in\Ph^+$. This is
called the canonical grading, and will correspond to the usual Toda theory
(termed abelian since $\cG_0$ is abelian in this case).
The general case of a non-abelian
$\cG_0$ will correspond to a non-abelian Toda theory.

We assume
henceforth that the Lie group $M_{\cG}$ whose Lie algebra is $\cG$ is
maximally non-compact.
We can then make a generalised Gauss decomposition of a group element
$g\in M_{\cG}$
according to the grading, writing
\eqn\eIviii{
g=g_-g_0g_+ }
where
\eqn\eIix{
g_-=\exp(\sum_{\ba\in\Ph^+\bs\tP}\ph^{\ba}_-E_{-\ba}),\qquad
 g_+=\exp(\sum_{\ba\in\Ph^+\bs\tP}\ph^{\ba}_+E_{\ba}),}
and where
\eqn\eIx{
g_0=\tg \exp(\sum_{\bi=r^{\tcG}+1}^{r^{\cG}}r^{\bi}H_{\bi}),\quad
\tg=\tg_-\tg_0\tg_+ }
with
\eqn\eIxi{\eqalign{
\tg_-&=\exp(\sum_{\ta\in\tP}\ph^{\ta}_-E_{-\ta}),\qquad
\tg_+=\exp(\sum_{\ta\in\tP}\ph^{\ta}_+E_{\ta}),\cr
\tg_0&=\exp(\sum_{\ti=1}^{r^{\tcG}}
r^{\ti}H_{\ti}). \cr}}
The parameters $\{\ph_-^{\ba},\ph_+^{\ba}\,(\ba\in\Ph^\bs\tP),\ph_-^{\ta},
\ph_+^{\ta}\,(\ta\in\tP),
r^i \, (i=1,\ldots r^{\cG})\}$ may be
regarded as co-ordinates on the group manifold $M_{\cG}$.

 We now write down the action for
the Wess-Zumino-Witten (WZW) model\ew\ defined on a group
manifold $M_{\cG}$:
\eqn\eIv{
kS_{WZW}(g)=-{k\over{8\p}}\int_S d^2x \tr(g^{-1}\del_{\m} gg^{-1}\del^{\m}g)
+{ik\over{12\p}}\int_B d^3x \e^{\m\n\r}\tr(g^{-1}\del_{\m} gg^{-1}\del_{\n}g
g^{-1}\del_{\r}g)  }
where $g\in M_{\cG}$, and where $B$ is a 3-dimensional ball whose
surface is the two-dimensional worldsheet $S$. We assume that the group
generators are normalised as in Eq. (A.5).
We are using here the conventions of Ref. \ref\go{P. Goddard and D. Olive,
\ijmpa1 (1986) 303.}. The level $x$ is defined in terms of $k$ by
$x={2k\over{\psi^2}}$, where $\psi$ is the highest root in
$\cG$. It is conventional to normalise so that $\psi^2=2$, in which case
$x=k$, but we cannot
in general normalise both $\cG$ and $\tcG$ in this fashion simultaneously, and
so we prefer to leave $\psi^2$ arbitrary. For a compact group $M_{\cG}$ with
Lie algebra $\cG$, $x$ is restricted to be an integer, but there is no such
constraint in the non-compact case considered here.
This action is invariant
under the transformations
\eqn\eIvi{
g(z,\bz)\rightarrow\O_L(z)g(z,\bz)\O_R(\bz), }
where we have introduced holomorphic and antiholomorphic co-ordinates
$z=x^0+ix^1, \quad \bz=x^0-ix^1$. The generators of the transformations
\eIvi\ are the currents
\eqn\eIvii{
J(z)=k\del gg^{-1},\qquad \bJ(\bz)=kg^{-1}\bdel g,}
which generate two commuting copies of the Kac-Moody algebra.
With a view to defining the non-abelian Toda theory, we first pick elements
$M_{\pm}$ in $\cG_{\pm}$ such that
\eqn\eIxii{
[\d.H,M_{\pm}]=\pm M_{\pm}. }
In general, any choice of $M_{\pm}$ satisfying Eq. \eIxii\ will suffice for
our purposes. However,
in order that the non-abelian Toda theory may provide a polynomial
realisation of the Kac-Moody algebra, one needs to be able to use the
Drinfeld-Sokolov gauge\ds. This is ensured if $M_-$ satisfies\FORTW
\eqn\eIxiia{
{\rm Ker}(\ad M_-)\cap\cG_+=\{0\}  }
and if, as we have arranged, the simple roots all have grade one or zero.
It appears\FORTW\ that there is one and only one possible such $M_-$ up to
conjugation by the little group of $\d.H$. Moreover, if Eq. \eIxiia\ is
satisfied, then one can find an $M_+$ such that $\{\d.H, M_{\pm}\}$
generate an $sl(2)$ embedding\FORTW,
i.e. in addition to Eq. \eIxii\ we also have
\eqn\eIxiib{
[M_+,M_-]=2\d.H.  }
However there is no need to use this particular $M_+$. (Conversely, given
an $sl(2)$ embedding into $\cG$ one can ask when one can find $\d$ and
$M_{\pm}$ satisfying Eqs. \eIxii\ and \eIxiia. This question is answered in
Ref. \FORTWi.)
The non-abelian Toda theory is now obtained from the WZW model in Eq. \eIv\ by
imposing the following constraints on the Kac-Moody currents in Eq. \eIvii\OW:
\eqn\eIxiii{
J_+=kM_+,\quad \bJ_-=kM_-,}
where $J_+$, $\bJ_-$ represent the projections of $J$, $\bJ$ onto $\cG_+$,
$\cG_-$ respectively.
The action $kS_{WZW}(g)$ in Eq. \eIvi\ now reduces classically to
\eqn\eIxiv{
S_{NAT}(g_0)=kS_{WZW}(g_0)-k\int d^2xV(\ph_-^{\ta},\ph_+^{\ta},r^i)}
where
\eqn\eIxiva{
V(\ph_-^{\ta},\ph_+^{\ta},r^i)={1\over{8\p }}\tr[M_+g_0M_-g_0^{-1}] }
and with $g_0$ as given by Eqs. \eIx, \eIxi. Eq. \eIxiv\ represents the action
for the non-abelian Toda theory. The kinetic part of the action consists
of a standard WZW action for the group $M_{\cG_0}$ whose Lie algebra is
$\cG_0$.  The remaining term in Eq.
\eIxiv\ yields a scalar potential term, typically consisting of a sum of
terms of the form $f(\ph^{\ta}_{\pm})e^{l(r^i)}$ where $f$, $l$ are
polynomials ($l$ is linear). Corresponding to the decomposition of $\cG_0$
into a direct sum of $\tcG$ together with factors of $R$,
the WZW action $S_{WZW}(g_0)$ may be written
\eqn\eIxv{
S_{WZW}(g_0)=S_{WZW}(\tg)-\sum_{\bi=r^{\tcG}+1}^{r^{\cG}}S_0(r^{\bi}) }
where $S_0(r)$ is the action for a free massless scalar field,
\eqn\eIxvi{
S_0(r)=\int d^2x\del_{\m}r\del^{\m}r }
and where $\tg$ is defined in Eqs. \eIx, \eIxi.

In the following Sections we shall determine the correct quantum form of
Eq. \eIxiv\ to ensure exact conformal invariance at the quantum level. This
will entail modifying the couplings and adding a dilaton field coupling to
the two-dimensional curvature.

\newsec{Conformal Invariance Conditions}
The original WZW model in Eq. \eIv\ is exactly conformally invariant at the
quantum level. In the standard language of conformal field theory, this means
that the energy-momentum tensor is traceless and its independent components
$T(z)$, $\bT(\bz)$ generate the Virasoro algebra. Moreover, for the WZW model
$T(z)$ and $\bT(\bz)$ can be written in terms of the Kac-Moody currents
in \eIvii\ according to the Sugawara construction
\ref\sug{H. Sugawara, \pr170 (1968) 1659.} (for a review see also \go) as
\eqn\eIIa{
T(z)={1\over{2k+c^{\cG}}}\tr(J^2),\quad
\bT(\bz)={1\over{2k+c^{\cG}}}\tr(\bJ^2), }
where $c^{\cG}$ is the value of the quadratic Casimir in the adjoint
representation of $\cG$, which is related to the dual Coxeter number $h^{\cG}$
by $h^{\cG}={c^{\cG}\over{\psi^2}}$.
The central charge is given by
\eqn\eIIaa{
c={k{\rm dim}\cG\over{k+{1\over2}c^{\cG}}}.  }
We wish to maintain
exact conformal invariance for the non-abelian Toda theory. The
energy-momentum tensor components of the reduced theory are given by\BFOFWII\OW
\eqn\eIIb{
T(z)={1\over{2k+c^{\cG}}}\tr(J^2)-\tr(\d.H\del J) }
(with a similar expression for $\bT(\bz)$), where the additional term
(with $\d$ given by Eq. \eIiii) is
required to ensure that the energy-momentum
tensor commutes with the constraints Eq. \eIxiii.

Our main purpose is to discuss the implications of the requirement of exact
conformal invariance for the precise form of the action for the non-abelian
Toda theory, which was given classically by Eq. \eIxiv. To this end it is
convenient to discuss conformal invariance in the non-linear $\s$-model
formulation. The action for a general non-linear $\s$-model may be written
\eqn\eIIc{
S(\ph)={\l\over{8\pi}}\int d^2x\{G_{ij}(\ph)\del_{\m}\ph^i\del^{\m}\ph^j
+\e^{\m\n}B_{ij}(\ph)\del_{\m}\ph^i\del_{\n}\ph^j
 +{1\over{\l}}D(\ph)R^{(2)}+V(\ph)\}  }
where $\e^{\m\n}$ is the two-dimensional alternating symbol,
$\{\ph^i\}$ represent co-ordinates on some target manifold with
metric $G_{ij}$ and antisymmetric tensor field $B_{ij}$ defined on it,
$D$ is the dilaton field coupling to the two-dimensional scalar curvature
$R^{(2)}$, and $V$ is the tachyon field. (The terminology derives from string
theory; our conventions are equivalent to taking $\a^{\pri}={2\over{\l}}$
in Ref. \ho, where $\a^{\pri}$ is the string coupling.)
The conformal invariance conditions for the $\s$-model Eq. \eIIc\ may
be written\betab\ho
\eqna\eIId{$$\eqalignno{
B^G_{ij}&\equiv\b_{ij}^G+{2\over{\l}}\nabla_i\del_jD+2\del_{(i}W_{j)}=0
&\eIId a\cr
B_{ij}^B&\equiv\b_{ij}^B+{2\over{\l}}H^k{}_{ij}\del_kD+2H^k{}_{ij}W_k=0
&\eIId b\cr
B^V&\equiv\b^V-2V+{1\over{\l}}\del^kD\del_kV+W^k\del_kV=0&\eIId c\cr}$$   }
where $\b^G_{ij}$, $\b^B_{ij}$ and $\b^V$ are the standard renormalisation
group $\b$-functions for $G_{ij}$, $B_{ij}$ and $V$, and $H_{ijk}$ is the
torsion, defined by $H_{ijk}=3\nabla_{[i}B_{jk]}$. $W_i$ is a vector field
which can be determined perturbatively within a given renormalisation
scheme. The results which we shall be using for the $\b$ functions imply
a renormalisation scheme in which $W_i$ vanishes.

When $B^G_{ij}$ and $B^B_{ij}$ both vanish, the quantity $B^D$ given by
\eqn\eIIe{
B^D\equiv\b^D+{1\over{\l}}\del^kD\del_kD+W^k\del_kD, }
where $\b^D$ is the dilaton $\b$-function, becomes constant
\ref\CP{G. Curci and G. Paffuti, \npb286 (1987) 399.}\ho\ and is then
related to the central charge $c$ for the conformal field theory by
\eqn\eIIf{
c=3B^D. }
In the case of the WZW model Eq. \eIv, the target manifold is the group
$M_{\cG}$.
There is no tachyon field $V$, and the
metric $G_{ij}$ and $B_{ij}$ may be read off by comparing Eqs. \eIv\ and
\eIIc. We have
\eqn\eIIfa{
G_{ij}=e_{ai}e_{aj}, \qquad \l=k, }
where the vielbein $e_{ai}$ is defined by
\eqn\eIIfb{
ig^{-1}\del_ig=e_{ia}T_a, }
with $T_a$ denoting a generic element of the Lie algebra, i.e. either
$E_{\pm\a}$ or $H_i$.
In fact $B_{ij}$ can only be defined locally, but we have
\eqn\eIIfc{
H_{ijk}={1\over2}f_{abc}e_{ai}e_{bj}e_{ck} }
These values for $G_{ij}$ and $H_{ijk}$
satisfy Eq. \eIId\ with vanishing $D$ and $W_i$, in virtue of the fact that
the generalised curvature given by
\eqn\eIIfd{
{\cal R}^i_{jkl}=R^i_{jkl}+2\del_{[l}H^i{}_{k]j}
+2H^m{}_{j[l}H^i{}_{k]m}}
vanishes, which is sufficient to imply the vanishing of $\b^G_{ij}$ and
$\b^B_{ij}$\ref\mukhi{S. Mukhi, \plb162 (1985) 345}.
The central charge given by Eq. \eIIf\ can be shown to reproduce the exact
value of Eq. \eIIaa\ at least to $O(k^{-2})$ in an
expansion in powers of ${1\over k}$
\ref\ajj{R. W. Allen, I. Jack and D. R. T. Jones, \zpc41 (1988) 323.} .

The question which now arises is the following: how can we perform the
reduction of the WZW action to the non-abelian Toda action in such a way as
to preserve the exact conformal invariance? The main problem is that the
non-abelian Toda action in Eq. \eIxiv\ contains a scalar potential and
hence $V$ in Eq. \eIIc\ is no longer zero, as was the case for the WZW model,
but rather is given by Eq. \eIxiva.
Hence we have to satisfy the additional conformal invariance condition Eq.
\eIId{c}\ in addition to Eqs. \eIId{a,b}. It is certainly not obvious at first
sight that Eq. \eIId{c}\ will be satisfied by the potential term which appears
in Eq. \eIxiva. The key is to identify Eq.
\eIId{c}\ with the condition for $V$ to be a primary field of conformal
dimensions $h=\bar h=1$ (in string theory terms, the condition for $V$ to be
a physical tachyon field). As such it should automatically be satisfied,
and we shall see later that indeed it is.
We can explicitly write down the corresponding Virasoro constraint on $V$
and by comparing with Eq. \eIId{c} we will be able to deduce the required
modifications of the coupling constants in Eq. \eIxiv. We will also find that
a non-zero dilaton field is now required.
\newsec{The Virasoro Constraint}
This section will be devoted to the discussion of the explicit form of the
Virasoro constraint on the potential $V$ in the non-abelian Toda theory.
The zero modes in Laurent expansions of the Kac-Moody currents $J$, $\bJ$
act as differential operators $\cJ^L$, $\cJ^R$
on functions of the co-ordinates of $M_{\cG}$, namely $\ph^{\a}_{\pm}$,
$\a\in\Ph^+$, $r^{\ta_i}$, $\ta_i\in\tD$, $r^i$, $i=r^{\tcG}+1,\ldots,r^{\cG}$.
The components of these operators, defined by
\eqn\eIIIa{
\cJ^L_a=\tr(T_a\cJ^L),\qquad \cJ^R_a=\tr(T_a\cJ^R), }
are in fact the generators of
left and right multiplication by Lie algebra elements, {\it viz.}
\eqn\eIIIb{
\cJ_a^Lg=T_ag, \qquad \cJ_a^Rg=gT_a. }
Mathematically, $\cJ^L$, $\cJ^R$ may be regarded as left- and right-invariant
vector fields respectively on the group manifold.
Correspondingly, the Virasoro generators corresponding to the energy-momentum
operator in \eIIb\ may be written as differential operators acting on
functions of the co-ordinates on the group manifold $M_{\cG}$. In particular,
the Virasoro operator $L_0$ can be expressed as a
differential operator by replacing the zero
modes of the currents in $T(z)$
in  Eq. \eIIb\ by $\cJ^L_a$, and similarly $\bar L_0$ may also be expressed
as an operator
by replacing the zero modes of the currents in $\bar T(\bz)$ by $\cJ^R_a$.
The results are
\eqn\eIIIc{\eqalign{
L_0&={1\over{2k+c^{\cG}}}\tr[(\cJ^L)^2]-\tr(\d.H\cJ^L), \cr
\bar L_0&={1\over{2k+c^{\cG}}}\tr[(\cJ^R)^2]-\tr(\d.H\cJ^R).  \cr}  }
The expressions  $\tr[(\cJ^L)^2]$ and $\tr[(\cJ^R)^2]$ are
in fact both equal to the Casimir operator $\cC^{\cG}$ for the group $\cG$.
 Moreover,
since $\d.H$ is in the Cartan subalgebra, $\tr(\d.H\cJ^L)
=\tr(\d.H\cJ^R)$.
Hence $L_0$ and $\bar L_0$ coincide as operators.
The Virasoro conditions for the potential
$V(\ph_-^{\ta},\ph_+^{\ta},r^i)$ in Eq. \eIxiva\ to be a primary field
of conformal dimensions $h=\bar h=1$ take the form
\eqn\eIIId{
(L_0+\bar L_0-2)V(\ph_-^{\ta},\ph_+^{\ta}, r^i)=0,\qquad
(L_0-\bar L_0)V(\ph_-^{\ta},\ph_+^{\ta},r^i)=0. }
The second of Eqs. \eIIId\ is automatically satisfied; the first becomes,
using Eq. \eIIIc,
\eqn\eIIIe{
({\cC^{\cG}\over{k+{1\over2}c^{\cG}}}-2\tr(\d.H\cJ^R)-2)V(\ph_-^{\ta},
\ph_+^{\ta},r^i)
=0. }
It thus becomes important to have some knowledge of the form of the
Casimir operator for co-ordinates corresponding to the generalised Gauss
decomposition of Eqs. \eIviii--\eIxi. This problem has been addressed in Ref.
\ref\ls{A. N. Leznov and M. V. Saveliev, Soviet J. Part. Nucl. {\bf 7}
(1976) 22; I. A. Fedoseev, A. N. Leznov and M. V. Saveliev,
\nc A76 (1983) 596.} and the results are summarised in
a recent book by Leznov and Saveliev
\ref\book{A. N. Leznov and M. V. Saveliev,
Group-theoretical methods for integration of non-linear dynamical
systems (Birkh\"auser, Basel, 1992).}. These results were restated and
amplified in Ref.\jp, to which we refer the reader for proofs of the
identities we shall use here.
It is convenient to
begin by introducing operators $X^L_{\pm\a}$, $X^R_{\pm\a}$, $\a\in\Ph^+$,
defined by
\eqn\eIIIf{\eqalign{
X_{-\a}^L(g_-\tg_-)=E_{-\a}(g_-\tg_-),
&\qquad X_{-\a}^R(g_-\tg_-)=(g_-\tg_-)E_{-\a}, \cr
X_{+\a}^L(\tg_+g_+)=E_{\a}(\tg_+g_+), &\qquad X_{+\a}^R(\tg_+g_+)=(\tg_+g_+)
E_{\a}. \cr}  }
The operators $X_{-\a}^{L,R}$, $\a\in\Ph^+$
act only on $g_-\tg_-$ and contain only the
variables $\ph_+^{\a}$, and similarly the operators $X_{+\a}^{L,R}$ act
only on $\tg_+g_+$ and contain only the variables $\ph_+^{\a}$. It can be
shown after some algebra that the Casimir operator $\cC^{\cG}$ is given by
\eqn\eIIIg{\eqalign{
\cC^{\cG}&=\sum_{\b\in\Ph^+}2e^{-r^i\b_i}X_{-\b}^RX_{+\b}^L
+2\r^i{\del\over{\del r^i}}\cr&\qquad
+{\del\over{\del r^i}}{\del\over{\del r^i}},
\cr}   }
where
\eqn\eIIIga{
\r={1\over2}\sum_{\a\in\cG}\a.  }
It is clear in fact that $X_{-\ta}^R$ acts only on $\tg_-$, and contains
only $\ph_-^{\ta}$, and $X_{+\ta}^L$ acts only on $\tg^+$, and contains
only $\ph_+^{\ta}$. They can be given explicitly by analogous expressions to
those obtained in Ref.\jp\  for the canonical decomposition; for instance,
\eqn\eIIIh{
X_{+\ta}^L=\sum_{n=0}^{\infty}{b_n\over{n!}}
N_{\a\tb_1\ldots\tb_n}\ph_+^{\tb_1}\ldots\ph_+^{\tb_n}
\del_{+(\a+\tb_1+\ldots
+\tb_n)} }
where $N_{\a\tb_1\ldots\tb_n}$ is the coefficient of
$E_{\a+\tb_1+\ldots
\tb_n}$ in $[E_{\tb_n},\ldots[E_{\tb_2},[E_{\tb_1},
E_{\a}]]\ldots]$, $b_n$ are the Bernoulli numbers, and
$\del_{\pm\a}={\del\over{\del\ph_{\pm}^{\a}}}$. $X^R_{-\ta}$ may be
obtained from Eq. \eIIIh\ by replacing the subscript $+$ by $-$ and
 inserting a factor $(-1)^n$ in the summation. The most
important property to notice is that $X^L_{+\ta}$ contains only
derivatives with respect to $\ph_+^{\tb}$ for $\tb\geq\ta$, and
$X^R_{-\ta}$ contains only
derivatives with respect to $\ph_-^{\tb}$ for $\tb\geq\ta$. In
particular, for the highest root $\tb\in\tP$,
\eqn\eIIIi{
X^L_{+\tb}={\del\over{\del\ph_+^{\tb}}},\quad
X^R_{-\tb}={\del\over{\del\ph_-^{\tb}}}.}
We will not give expressions for $X_{\pm\ba}^{L,R}$ here, as they are even
more complicated and will not be required in any case. It suffices to note
that they have the property just mentioned, namely that $X^L_{+\ba}$
contains only derivatives with respect to $\ph_+^{\bar \b}$ for
$\bar\b\ge\ba$, and similarly for $X_{-\ba}^R$. Finally it is
straightforward to show using the methods of Ref. \jp\ that
\eqn\eIIIia{\eqalign{
\cJ_{\ti}^R&={\del\over{\del r^{\ti}}}-\sum_{\ta\in\tP}\ph_+^{\ta}
\ta_{\ti}\del_{+\ta}+\ldots\cr
\cJ_{\bi}^R&={\del\over{\del r^{\bi}}}+\ldots\cr} }
where, in accord with the notation of Eq. \eIIIb, $\cJ_{\ti}^R$ is the
operator inducing multiplication of $g$ on the right by $H_{\ti}$. We have
omitted terms involving derivatives with respect to $\ph_+^{\ba},\quad\ba\in
\Ph^+\bs\tP$, since these do not contribute when acting on
$V(\ph_-^{\ta},\ph_+^{\ta},r^i)$. There are similar expressions for $\cJ_{\ti}
^L$, $\cJ_{\bi}^L$. Using the fact that $\d.\ta=0$ for $\ta\in\tP$, which
follows from \eIiii, we have
\eqn\eIIIj{
\tr(\d.H\cJ^R)=\tr(\d.H\cJ^L)=\d_i{\del\over{\del r^i}}=
\d_{\bi}{\del\over{\del r^{\bi}}}.    }
Using Eqs. \eIIIg, \eIIIj, we can now write the Virasoro condition Eq.
\eIIIe\ explicitly. Moreover, since $V$ is independent of $\ph_{\pm}^{\ba}$,
the operators $X_{\pm\ba}^{L,R}$ can be omitted from
the expression for $\cC^{\cG}$ in Eq. \eIIIg\ for our
purposes.

It is possible to check explicitly that the Virasoro condition Eq. \eIIIe\ is
satisfied, with $V(\ph_-^{\ta},\ph_+^{\ta},r^i)$ as given by Eq. \eIxiva.
In fact, we have the stronger results
\eqna\eIIIk$$\eqalignno{
\cC^{\cG} V(\ph_-^{\ta},\ph_+^{\ta},r^i)&=0,&\eIIIk a  \cr
(\tr(\d.H\cJ_R)+1)V(\ph_-^{\ta},\ph_+^{\ta},r^i)&=0.&\eIIIk b \cr}  $$
These identities provide a concise statement of the properties
of the potential $V$ which guarantee conformal invariance at the quantum level.
The proof of Eq. \eIIIk{a}
is straightforward but quite lengthy, and we relegate it to the
Appendix; Eq. \eIIIk{b}\ follows straightforwardly from Eqs. \eIix,
\eIxii, \eIxiva\ and \eIIIj.
\newsec{The tachyon $\b$-function}
In this Section
we obtain a more explicit form for the tachyon $\b$-function with
a view to comparing it with the explicit form of the Virasoro constraint for
$V$ obtained in the previous Section. This will enable us to determine the
modified form of the non-abelian Toda action in Eq. \eIxiv\ which will
ensure full conformal invariance. We postulate that the correct action for
the non-abelian Toda theory is given by
\eqn\eIVa{
S_{NAT}(g_0)=k^{\pri}S_{WZW}(\tg)-k^{\pp}\sum_{i=r^{\tcG}+1}^{r^{\cG}}S_0(r^i)+
\m \int d^2x V(\ph_-^{\ta},\ph_+^{\ta},r^i)
+\int d^2x D(r^{\ta_i},r^i)R^{(2)}.  }
where $V(\ph_-^{\ta},\ph_+^{\ta},r^i)$ is as defined in Eq. \eIxiva, and $\m$
is a constant (which we will not determine).
We wish to discover the form of $D$ and to obtain $k^{\pri}$, $k^{\pp}$ in
terms
 of the
coupling $k$ of the original WZW model in Eq. \eIv. Before proceeding further,
we need a more explicit form for the tachyon $\b$-function which appears in
Eq. \eIId{c}. For a general $\s$-model given by Eq. \eIIc, the tachyon
$\b$-function can be calculated perturbatively, and takes the form
\betaa\ref\bos{M. Bos, \apny181 (1988) 177.}
\ref\hoa{H. Osborn, \apny200 (1990) 1.}
\eqn\eIVc{
\b^V=[-{1\over{\l}}\nabla^2+
2{1\over{\l^2}}H^i{}_{kl}H^{jkl}\nabla_i\nabla_j+\ldots]V}
However, to attain our goal of exact conformal invariance we need an exact
expression for the tachyon $\b$-function. It can be seen by
power counting that the tachyon $\b$-function must take the form
\eqn\eIVd{
\b^V=\O V }
with $\O$ depending only on the metric $G_{ij}$ and the antisymmetric
tensor field $B_{ij}$ in Eq. \eIIc\ (i.e. not on $D$ or $V$).
The exact form of $\O$ is not known in the general
case, but it will be sufficient to have the exact result for the particular
instance of the WZW model. This will enable us to compute the exact
form for $\O$  for the action in Eq. \eIVa, since
the value of $\O$ for a sum of kinetic terms such as
appears in Eq. \eIVa\ is the sum of the values of $\O$
for each individual kinetic term, and the value of $\O$ for the free field
kinetic action is
trivial (simply $-\del^2$). For the WZW model of
Eq. \eIv, the calculation of $\O$ is essentially the same as the
calculation of the anomalous dimension of $g$\bos, and hence the exact result
for $\O$ can be deduced as\ref\jj{I. Jack and D. R. T. Jones, in
preparation.}
\eqn\eIVe{
\O=-{1\over{k+{1\over2}c^{\cG}}}\nabla^2.  }
The perturbative result given in Eq. \eIVc\ is readily seen to agree with
Eq. \eIVe\ upon use of Eqs. \eIIfa, \eIIfc, with $c^{\cG}\d_{ab}=f_{acd}
f_{bcd}$.
It will be convenient from our point of view to write $\nabla^2$ in the
equivalent form
\eqn\eIVf{
\nabla^2={1\over{\sqrt G}}\del_i\sqrt G G^{ij}\del_j ,}
where $G= |\det (G_{ij})|$.
Finally we can identify the conformal invariance condition for the tachyon,
Eq. \eIId{c}, with the Virasoro condition on $V$, Eq. \eIIIe. Denoting the
co-ordinates of $\tcG$ generically by $x^{\tra}$, and the corresponding metric
for $S_{WZW}(\tg)$ by $\tG_{\tra\trb}$, we have
\eqn\eIVg{\eqalign{
-{1\over{k^{\pri}+{1\over2}c^{\tcG}}}{1\over{\sqrt {\tG}}}\del_{\tra}\sqrt{
\tG}
\tG^{\tra\trb}\del_{\trb} +{1\over{ k^{\pp}}}\sum_{\bi=r^{\tcG}+1}^{r^{\tG}}
{\del^2\over{\del
r^{\bi2}}}&+{1\over{k^{\pri}}}\del_{\tra}DG^{\tra\trb}\del_{\trb}
-{1\over{ k^{\pp}}}\sum_{\bi=r^{\tcG}+1}^{r^{\tG}}\del_{\bi} D
\del_{\bi}\cr
=\bigl({\cC^{\cG}\over{k+{1\over2}c^{\cG}}}-2(\d.H,\cJ_R)&\bigr) \cr} }
where $\del_i\equiv{\del\over{\del r^i}}$, etc, and where
\eqn\eIVh{\eqalign{
\cC^{\cG}&=\sum_{\tb\in\tP}2e^{-r^{\ti}\tb_{\ti}}X_{-\tb}^RX_{+\tb}^L
+2\r_i{\del\over{\del r^i}}\cr&\qquad
+{\del\over{\del r^i}}{\del\over{\del r^i}}
\cr}   }
as in Eq. \eIIIg, but recalling that the operators $X_{\pm\ba}^{L,R}$ can be
omitted from $\cC^{\cG}$ when acting on $V(\ph_-^{\ta},\ph_+^{\ta},r^i)$. The
metric $G_{\tra\trb}$, the dilaton and the couplings $k^{\pri}$, $k^{\pp}$
can now be read off by comparing the LHS and RHS of Eq. \eIVg. Firstly, we
manifestly must have
\eqn\eIVi{
k^{\pp}=\k }
where for convenience we define
\eqn\eIVj{
\k=k+{1\over2}c^{\cG}. }
Moreover, the metric $G_{\tra\trb}$ can be read off by comparing the double
derivatives with respect to $\ph_{\pm}^{\ta}$, $r^{\ti}$ on the LHS and
RHS of Eq. \eIVg. These terms in the Casimir Eq. \eIVh\ are exactly the same
as they would appear in the Casimir for $\tcG$, and so the metric
$G_{\tra\trb}$ is proportional to the standard metric for the WZW model
corresponding to $\cG$. The normalisation is fixed by requiring that
classically $k^{\pri}=k$, which implies
\eqn\eIVk{
k^{\pri}=\k-{1\over2}c^{\tcG}.   }
In particular, we now have
\eqn\eIVka{
G^{\ti\tj}=-\d^{\ti\tj}. }
Using the property of $X_{+\ta}^L$ and $X_{-\ta}^R$ mentioned in Section 4,
namely that
they contain only derivatives with respect to $\ph_{\pm}^{\b}$  respectively
for $\b>\ta$, we can show using row and column operations that
\eqn\eIVl{
G= |\det G|=\exp(2r_{\ti}\trho_{\ti}),  }
where
\eqn\eIVla{
\trho={1\over2}\sum_{\a\in\tP}\ta .  }
The form of $X_{+\ta}^R$ given in Eq. \eIIIh\ (with the similar form
for $X_{-\ta}^R$) also implies a form for the metric $G^{\tra,\trb}$ which
guarantees that
${1\over{\sqrt {\tG}}}\del_{\tra}\sqrt{ \tG}
\tG^{\tra\trb}\del_{\trb}$ contains no contributions linear in derivatives
${\del\over{\del \ph_{\pm}^{\ta}}}$. Hence the only terms linear in
derivatives on the LHS of Eq. \eIVg\ come from ${1\over{k^{\pp}+{1\over2}
c^{\tcG}}}
{1\over{\sqrt G}}\del_{\ti}\sqrt G G^{\ti\tj}
\del_{\tj}$. Using Eqs. \eIVka, \eIVk\ and \eIVl, we find
\eqn\eIVm{
{1\over{k^{\pri}+{1\over2}
c^{\tcG}}}
{1\over{\sqrt G}}\del_{\ti}\sqrt G G^{\ti\tj}
\del_{\tj}={1\over{\k}}\del_{\ti}\del_{\ti}
+{2\over{\k}}\trho_{\ti}\del_{\ti}.  }
Comparing the
terms first order in derivatives on the LHS and RHS of Eq. \eIVg, we find
using Eqs. \eIii\ and \eIiii
\eqn\eIVn{
{2\over{\k}}(\r_{\bi}\del_{\bi}+\r_{\pri {\ti}}\del_{\ti})+{1\over2}\d_i\del_i
=-{1\over{k^{\pri}}}\del_{\ti}D\del_{\ti}
-{1\over{k^{\pp}}}\del_{\bi}D\del_{\bi}, }
where
\eqn\eIVna{
\r^{\pri}=\r-\trho,  }
with $\r$, $\trho$ as given in Eqs. \eIIIga, \eIVla.
We show in Appendix A that
\eqn\eIVo{
\r^{\pri}.\ta=0,\qquad\ta\in\tP, }
from which it follows, using Eq. \eIiva that
\eqn\eIVp{
\r^{\pri}_{\ti}=0.  }
Similarly, it follows that
\eqn\eIVq{
\d_{\ti}=0. }
We finally deduce
\eqn\eIVr{
D=-2\r_{\bi}r^{\bi}-{1\over2}
\k\d_{\bi}r^{\bi}, }
which, using Eqs. \eIVna, \eIVp, \eIVq\ and the fact that $\trho_{\bi}=0$,
we may write in the form
\eqn\eIVra{
D=-2\r^{\pri}_ir_i-\k\d_ir^i. }
It is in fact essential that $D$ should not depend on $r^{\ti}$
(or indeed any of the variables associated with $\tcG$):
The metric, anti-symmetric tensor and dilaton should satisfy all the
conformal invariance conditions Eq. \eIId{}. The metric and antisymmetric
tensor field $G_{\tra\trb}$ and $B_{\tra\trb}$ are exactly those corresponding
to the WZW model for $\tcG$. They satisfy Eq. \eIId{a}\ without the need
for a dilaton field, in other words $\b^G_{\tra\trb}$ and
$\b^B_{\tra\trb}$ vanish. The
Christoffel symbols $\G_{\tra\trb}^{\tilde c}$
are in general non-zero, and so if $D$
had any dependence on $r^{\ti}$ or $\ph_{\pm}^{\ta}$
then the RHS of Eq. \eIId{a}\ could not
vanish. (We could not remedy this by invoking a non-zero $W_i$ since
$D$ and $W_i$ appear in each of Eqs. \eIId{}\ in the combination
$\del_i D+2W_i$, and hence it is really this quantity which we identify in
Eq. \eIVr. We choose however to assume a renormalisation prescription for
which $W_i$ vanishes.) It is thus a good check on our method that $D$ only
depends on $r^{\bi}$. As a second, even more stringent check on our results we
shall now compute the central charge, obtaining agreement with previous
results.
The central charge is given by Eqs. \eIIe, \eIIf. The contributions to $c$
from $\b^D$ in Eq. \eIIe\ consist of the sum of the central charge for
the $(r^{\cG}-r^{\tcG})$ scalars $r^{\bi}$, and the central charge for the WZW
model corresponding to $\tcG$. So we have, using Eq. \eIVra,
\eqn\eIVs{
c=r^{\cG}-r^{\tcG}+{k^{\pri}{\rm dim}\tcG\over{k^{\pri}+{1\over2}c^{\tcG}}}
-{12\over{\k}}(\r^{\pri}+\k\d)^2,   }
which becomes, using Eqs. \eIiiia, \eIVk, \eIVna  and \eIVo,
\eqn\eIVt{\eqalign{
c&=r^{\cG}-r^{\tcG}+{1\over{\k}}{\rm dim}\tcG(\k-{1\over2}c^{\tcG})\cr
&\quad +{12\over{\k}}\trho^2 \cr
&\quad-{12\over{\k}}(\r+\k\d)^2. \cr} }
Using the Freudenthal-deVries strange formula this expression simplifies to
\eqn\eIVu{
c={\rm dim}\cG_0+{12\over{\k}}(\r+\k\d)^2, }
where
\eqn\eIVv{
{\rm dim}\cG_0=r^{\cG}-r^{\tcG}+{\rm dim}\tcG }
is the dimension of $\cG_0$.
Eq. \eIVu\ is the formula for the central charge obtained previously by other
methods\OW, generalising the result for the abelian Toda field theory
computed in Ref. \pma.
\newsec{An explicit example}
Our discussion so far has been somewhat abstract, and so it seems appropriate
to present a simple example for which we can give explicit expressions for all
the quantities involved. We take the case of the algebra $B_2$, which
is also the case discussed in Ref.\gs. The Cartan
matrix is given by
\eqn\Va{
A=\left(\matrix{2&-2\cr
                -1&2\cr}\right). }
Denoting the simple roots by two-vectors $\ua_1$, $\ua_2$, the
positive roots are
$\Ph^+=\{\ua_1,\ua_2,\ua_3=\ua_1+\ua_2,\ua_4=\ua_1+2\ua_2\}$.
The corresponding generators can be written in a $4\times4$ matrix
representation as
\eqn\Vb{\eqalign{
E_{\ua_1}=\left(\matrix{0&0&0&0\cr
                 0&0&1&0\cr
                 0&0&0&0\cr
                 0&0&0&0\cr}\right),&\qquad
E_{\ua_2}={1\over{\sqrt 2}}\left(\matrix{0&1&0&0\cr
                  0&0&0&0\cr
                  0&0&0&-1\cr
                  0&0&0&0\cr}\right),\cr
E_{\ua_3}={1\over{\sqrt 2}}\left(\matrix{0&0&1&0\cr
                  0&0&0&1\cr
                  0&0&0&0\cr
                  0&0&0&0\cr}\right),&\qquad
E_{\ua_4}=\left(\matrix{0&0&0&1\cr
                  0&0&0&0\cr
                  0&0&0&0\cr
                  0&0&0&0\cr}\right),\cr } }
with the generators corresponding to the negative roots, which we shall
denote as $E_{\ua_{-1}}$, $E_{\ua_{-2}}$, $E_{\ua_{-3}}$, and $E_{\ua_{-4}}$,
given by the
transposes of $E_{\ua_1}$--$E_{\ua_4}$. We pick the Cartan subalgebra
generators to be
\eqn\eVc{
H_1={1\over{\sqrt 2}}\left(\matrix{1&0&0&0\cr
                  0&0&0&0\cr
                  0&0&0&0\cr
                  0&0&0&-1\cr}\right),\qquad
H_2={1\over{\sqrt 2}}\left(\matrix{0&0&0&0\cr
                  0&1&0&0\cr
                  0&0&-1&0\cr
                  0&0&0&0\cr}\right).}
The generators have been normalised according to Eq. (A.5).
The commutation relations take the form
\eqn\eVe{
[ \underline H,E_{\ua_i}]=\underline{\a}_iE_{\ua_i} }
where
\eqn\eVf{
\underline{\a}_1={1\over{\sqrt 2}}\left(\matrix{0\cr2\cr}\right),\quad
\underline{\a}_2={1\over{\sqrt 2}}\left(\matrix{1\cr-1\cr}\right),\quad
\underline{\a}_3={1\over{\sqrt 2}}\left(\matrix{1\cr1\cr}\right),\quad
\underline{\a}_4={1\over{\sqrt 2}}\left(\matrix{2\cr0\cr}\right). }
so that we have
\eqn\eVg{
\ua_1^2=\ua_4^2=2,\qquad \ua_2^2=\ua_3^2=1.  }
We now pick $\tD$ to consist of the single simple root $\ua_2$. We
then find from Eqs. \eIii, \eIiii,
\eqn\eVh{
\ud=\ua_1+\ua_2=\a_3   }
so that
\eqn\eVi{
\underline{\d}.\underline H={1\over{\sqrt 2}}(H_1+H_2). }
It is then easy to check explicitly that
\eqn\eVj{
\ud.\ua_1=1, \quad \ud.\ua_2=0, \quad [\ud.\uH,E_{\ua_1}]=E_{\ua_1},
\quad[\ud.\uH,E_{\ua_2}]=0, }
as in Eqs. \eIiiia, \eIiv. We now parametrise a group element $g$ as in Eqs.
\eIviii--\eIxi, with
\eqn\eVk{\eqalign{
g&=g_-g_0g_+,\qquad g_0=\tg e^{s\ud.\uH},\qquad \tg=\tg_-\tg_0\tg_+,\qquad
\tg_0=e^{r\ua_2.\uH}\cr
g_-&=e^{(\ph^-_1E_{-\ua_1}+\ph^-_3E_{-\ua_3}+\ph^-_4E_{-\ua_4})},\qquad
\tg_-=e^{\ph^-_2E_{-\ua_2}},\cr
g_+&=e^{(\ph^+_1E_{\ua_1}+\ph^+_3E_{\ua_3}+\ph^+_4E_{\ua_4})},\qquad
\tg_+=e^{\ph^+_2E_{\ua_2}},\cr  }  }
where $\{\ua_2.\uH,\ud.\uH\}$ form an orthogonal basis for the Cartan
subalgebra
constructed as specified in Section 2.

We finally need to pick the elements $M_{\pm}$.
{}From the considerations presented in the Appendix, the simplest choice
would be $M_{\pm}=E_{\pm\ua_1}$. However, this leads to a fairly
trivial potential depending only on $r$ and $s$ in Eq. \eVk.
This is an example of a general phenomenon: if we take $M_{\pm}=
E_{\pm\bar{\g}_L}$, where $\bar{\g}_L$ is a simple root in $\D\bs\tD$, then
the potential  $V(\ph_-^{\ta},\ph_+^{\ta},r^i)$ depends only on the variables
$r^i$, and is in fact simply the potential for the corresponding
ordinary (i.e. abelian) Toda theory. This is because
\eqn\eVl{
\tg_-^{-1}E_{\bar \g_L}\tg_-=E_{\bar\g_L},\qquad
\tg_+E_{-\bar \g_L}\tg_+^{-1}=E_{-\bar\g_L} }
for $\bar\g_L\in\D\bs\tD$. Hence to make things more interesting, we pick
\eqn\eVla{
M_{\pm}=E_{\pm\ua_3}, }
which, since $\ua_3=\ua_1+\ua_2$,
in view of Eq. \eVj\ also have the properties
Eq. \eIxii. This $M_-$ also satisfies Eq. \eIxiia, guaranteeing the
existence of the Drinfeld-Sokolov gauge. Hence,
in this case $\{M_{\pm},\d.H\}=\{E_{\pm\ua_3},\ua_3.\uH\}$
generate a non-canonical embedding of $A_1$ in $B_2$.
We can now calculate the classical action as given by Eq. \eIxiv,
\eIxv.
The easiest way to do this is to use the Polyakov-Wiegmann identity
\eqn\eVm{\eqalign{
S_{WZW}(\tg_0)&=S_{WZW}(\tg_-)+S_{WZW}(\tg_0)+S_{WZW}(\tg_+)\cr
&\quad -{1\over{4\p}}\int d^2x\tr[(\tg_-^{-1}\bdel\tg_-)(\del\tg_0\tg_0^{-1}
+(\tg_0^{-1}\bdel\tg_0)(\del\tg_+\tg_+^{-1})\cr&\quad
+(\tg_-^{-1}\bdel\tg_-)\tg_0(\del\tg_+\tg_+^{-1})\tg_0^{-1}],\cr }}
with $g_0$, $\tg_{\pm}$ and $\tg_0$ as given by Eq. \eVk. Using Eq. (A.5), we
obtain
\eqn\eVn{
S_{NAT}(g_0)=-{k\over{8\p}}\int d^2x[\del r\bdel r
+\del s\bdel s+2e^{-2r}\del\ph_2^+
\bdel\ph_2^-  +V(\ph_2^{\pm},r,s)], }
where
\eqn\eVo{
V(\ph_2^{\pm},r,s)=e^{-s}+\ph_2^+\ph_2^-e^{r-s}.}
This action is of the form \eIxv\ with $s$ as the free scalar field.
We may write it in the form Eq. \eIIc\ with a metric $G_{ij}$ and an
antisymmetric tensor field $B_{ij}$ given by
\eqn\eVp{\eqalign{
G_{ij}&=\bordermatrix{&s&\tra\cr
                     s&-1&0\cr
                     \trb&0&\tG_{\tra\trb}\cr},\cr
\tG_{rr}&=-1,\qquad \tG_{+-}=\tG_{-+}=-e^{r},\cr
\tG^{rr}&=-1,\qquad \tG^{+-}=\tG^{-+}=-e^{r},\cr
B_{+-}&=\tB_{+-}=-B_{-+}=-\tB_{+-}=ie^{r},\cr}}
all other components being zero (with $\pm$ denoting the components
corresponding to $\ph_2^{\pm}$). $\tG_{ij}$ and $\tB_{ij}$ are then the metric
and antisymmetric tensor field for $S_{WZW}(\tg)$.
The Christoffel symbols and torsion are then
given by
\eqn\eVq{
\G^r_{+-}=-{1\over2}e^{r},\quad\G^+_{+r}=\G^-_{-r}={1\over2},\quad H_{+-r}=
{i\over2}e^{r}, }
with all other components not related by symmetry vanishing. We then
readily check that the generalised curvature given by Eq. \eIIfd\
vanishes, which is sufficient to ensure that the $\b$-functions $\b^G_{ij}$
and $\b^B_{ij}$ vanish. This is of course a consequence of the fact that
the kinetic part of the action Eq. \eVn\ is simply the that for the WZW model
for $A_1$ coupled to a free scalar. Hence Eqs. \eIId{a,b} are satisfied
without the dilaton $D$. We now wish to determine $D$ and the modified
couplings $k^{\pri}$ and $k^{\pp}$ in Eq. \eIVa\ using Eq. \eIVg, and we
shall see that the form for $D$ which emerges will be such as not to contribute
to Eqs. \eIId{a,b}. Moreover, once we have made the identification Eq. \eIVg,
the third conformal invariance condition Eq. \eIId{c}\ is guaranteed to be
satisfied; the LHS of Eq. \eIVg\ is the operator which acts on the potential
$V$ in Eq. \eIId{c}, and the RHS annihilates the potential
$V(\ph_2^{\pm},r,s)$ in Eq. \eVn, as we shall now demonstrate explicitly. We
first of all need an explicit form for the Casimir in Eq. \eIIIg.
We find for the operators $X_{+\a}^L$ defined in Eq. \eIIIf:
\eqn\eVt{\eqalign{
X_{+\ua_1}^L={\del\over{\del\ph_1^+}}-\ph_2^+{\del\over{\del\ph_3^+}}
+{1\over2}\ph_2^{+2}{\del\over{\del\ph_4^+}},&\quad X_{+\ua_2}^L=
{\del\over{\del\ph_2^+}},\cr
X_{+\ua_3}^L={\del\over{\del\ph_3^+}}-\ph_2^+{\del\over{\del\ph_4^+}},&\quad
X_{+\ua_4}^L={\del\over{\del\ph_4^+}},\cr}}
the operators $X_{-\ua_i}^R$ being given by the same expressions but with
$+\rightarrow-$. (These expressions were calculated using the algebraic
manipulation package REDUCE.)
The Casimir $\cC^{\cG}$ is now given by Eq. \eIIIg,
with Eqs. \eIIIga, \eVf, as
\eqn\eVu{\eqalign{
\cC^{\cG}&=2(e^{r-s}X_{-\ua_1}^RX_{+\ua_1}^L+e^{-r}X_{-\ua_2}^RX_{+\ua_2}^L
+e^{-s}X_{-\ua_3}^RX_{+\ua_3}^L+e^{-r-s}X_{-\ua_4}^RX_{+\ua_4}^L)\cr
&\quad +{\del^2\over{\del r^2}}+{\del^2\over{\del s^2}}+{\del\over{\del r}}
+3{\del\over{\del s}}.\cr}}
It is then easy to check that
\eqn\eVv{
\cC^{\cG}V(\ph_2^{\pm},r,s)=\bigr(2e^{-r}{\del\over{\del\ph^+_2\del\ph^-_2}}
+{\del^2\over{\del r^2}}+{\del^2\over{\del s^2}}+{\del\over{\del r}}
+3{\del\over{\del s}}\bigl)V(\ph_2^{\pm},r,s)=0,}
with $V(\ph_2^{\pm},r,s)$ as given by Eq. \eVo. We also have, from
Eq. \eIIIj\ and \eVh,
\eqn\eVw{
\tr(\ud.\uH\cJ^R)={\del\over{\del s}}}
from which it follows immediately that
\eqn\eVx{
(\tr(\ud.\uH\cJ^R)+1)V(\ph_2^{\pm},r,s)=0, }
and hence from Eqs. \eVv\ and \eVx, we see that indeed the RHS of Eq. \eIVg\
annihilates $V(\ph_2^{\pm},r,s)$ in this example. We must now identify the
quantities on the LHS of Eq. \eIVg. With $\tG_{\tra\trb}$ given by Eq. \eVp, we
have
\eqn\eVy{
\sqrt {\tG}=e^r,}
and so the LHS of Eq. \eIVg\ becomes
\eqn\eVz{
{1\over{k^{\pri}+1}}\bigl(2e^{-r}{\del\over{\del\ph^+_2\del\ph^-_2}}
+e^{-r}{\del\over{\del r}}e^{r}{\del\over{\del r}}+{1\over{k^{\pp}}}
{\del^2\over{\del s^2}}\bigr)
+({1\over{k^{\pri}}}{\del\over{\del r}})D{\del\over{\del r}}
+\bigl({1\over{k^{\pp}}}{\del\over{\del s}}D\bigr){\del\over{\del s}},  }
incorporating the obvious fact
that $D$ cannot depend on $\ph^{\pm}_2$, since no
 single derivatives with respect to $\ph^{\pm}_2$ appear. The RHS of Eq.
\eIVg\ as given explicitly by Eqs. \eVv\ and  \eVw\ takes the form
\eqn\eVza{ {1\over{\k}}\bigr(2e^{-r}{\del\over{\del\ph^+_2\del\ph^-_2}}
+{\del^2\over{\del r^2}}+{\del^2\over{\del s^2}}+{\del\over{\del r}}
+3{\del\over{\del s}}\bigl)+{\del\over{\del s}}+1,}
where, since
$c^{\cG}=3$ for $B_2$, $\k=k+{3\over2}$. Comparing \eVz\ and \eVza,
we must take
\eqn\eVzb{
k^{\pri}=\k-1,\qquad k^{\pp}=\k,\qquad
D=(3+\k)s,\qquad}
which agrees with the general result Eq. \eIVra,
since, from Eqs. \eIVna, \eVf\ and \eVh, we have
\eqn\eVzc{
\ur^{\pri}={3\over{\sqrt 2}}\left(\matrix{1\cr1\cr}\right)=3\ud.} 
emphasised in discussing the general case, the expression for $D$ in Eq,
\eVzb\ has the vital property that it does not depend on the variables
associated with $\tcG$. For instance, had the dilaton depended on $r$,
then from Eqs. \eIId{a,b}, \eVq, there would have been contributions
from the dilaton to $B^G_{ij}$, $B^B_{ij}$, which would therefore not
vanish (since $\b^G_{ij}$, $\b^B_{ij}$ are zero).
\newsec{Conclusions}
We have demonstrated how the process of hamiltonian reduction of the WZW
model may be implemented at the quantum level, to furnish an action for
the  non-abelian Toda theory which is exactly conformally invariant. The
quantum action differs from the classical version by the adjustment of
the  coupling constants and by the addition of a dilaton field. The
scalar  potential $V(\ph_-^{\ta},\ph_+^{\ta},r^i)$ satisfies the simple
identities \eIIIk{}\ which guarantee that it satisfies the Virasoro
constraint (at zeroth order in the Laurent expansion in $z$). These
identities are proved in the Appendix; the first of them follows from a
simple identity Eq. (A.16) for the ordinary  Lie algebra Casimir (i.e. not the
operator form of the Casimir), for the  Lie subalgebra $\cG_0$ which is
the grade-zero part of $\cG$.

The exact results for string black hole solutions, which were first obtained
using methods similar to those used here, have recently been rederived using
an approach based on the exact quantum effective action for the gauged WZW
model
\ref\aatI{A. A. Tseytlin, \npb399 (1993) 601; ``Conformal
sigma models corresponding to gauged Wess-Zumino-Witten theories'',
preprint CERN-TH.6804/93 (hep-th/9302083)\semi
I. Bars and K. Sfetsos, \prd48 (1993) 844}. (See also Ref. \ref\TS{
A. Giveon, E. Rabinovici and A. A. Tseytlin, ``Heterotic string solutions
and coset conformal field theories'', preprint CERN-TH.6872/93
(hep-th/9304155)\semi
K. Sfetsos and A. A. Tseytlin, ``Chiral gauged WZNW models and heterotic
string backgrounds'', preprint CERN-TH.6962/93 (hep-th/9308018).}.
This method presents conceptual advantages over the present technique
and moreover can also yield exact results for the antisymmetric tensor field,
if present. The antisymmetric tensor field cannot be treated by the methods
used here since it does not contribute to $L_0$, the Virasoro operator.
It would be desirable to extend the quantum effective action methods to the
present Toda field theory case.

Our initial interest in non-abelian Toda field theories was awakened by the
fact\gs that the kinetic sector of the non-abelian Toda theory based on $B_2$
was equivalent to Witten's string black hole\sbh. However, this has only
been shown at the classical level at present, and indeed the proof proceeds
by eliminating one of the fields using its equation of motion. It is therefore
not clear how one would demonstrate this equivalence at the quantum level. At
the most naive level, the equations of motion derived from the quantum
action we derived in Section 5 do not appear to produce the equations of
motion derived from the exact quantum action for the string black
hole\dvv\aatI\ even if we again allow ourselves to eliminate one of the fields
using its equation of motion. This point deserves more careful investigation,
though.
\vskip 12pt
\line{\bf Acknowledgements}
\vskip 5pt
I.J. thanks David Olive for a useful discussion. I.J. and J.P. thank the
S.E.R.C. for support.
\vskip 12pt
\appendix{A}{}
In this Appendix we summarise our conventions for Lie algebras, and also prove
some mathematical results, in particular Eqs. \eIIIk{}\ and \eIVo.
In general a Lie algebra $\cG$ can be specified by a
Cartan subalgebra $\cH$
of mutually commuting generators with a basis $\{H_i,\quad
i=1,\ldots r^{\cG}\}$, where $r^{\cG}$ is the rank of $\cG$, together with a
set of step operators $E_{\a}$ corresponding to the roots $\a$, satisfying
\eqn\Aaa{
[H_i,H_j]=0, \qquad [H_i,E_{\a}]=\a(H_i)E_{\a}, }
regarding the roots as elements of $\cH^*$, i.e. as maps from $\cH$ to $R$.
In terms of some ordered basis $\g_1,\ldots\g_{\r^{\cG}}$
for the roots, a root $\b$ can be defined as
positive if the first coefficient in the expansion of $\b$ in terms of the
$\g_i$ is positive. A simple root is a positive root
which cannot be written as the sum of two other positive roots.
The Cartan-Killing form is defined for two generators $X$, $Y$ of $\cG$ by
\eqn\Aab{
(X,Y)=\tr(\ad X\ad Y), }
where
\eqn\Aac{
(\ad X)Z=[X,Z].  }
We define
\eqn\Aad{
\g_{ij}=(H_i,H_j), \qquad \g_{\a\b}=(E_{\a},E_{\b}). }
All this is perfectly general; however, it is convenient to consider the case
of orthonormal generators, normalised so that
\eqn\Aae{
\tr(H_iH_j)=\d_{ij}, \qquad \tr(E_{-\a}E_{\b})=\d_{\a\b}. }
In these circumstances we have
\eqn\Aaea{
\g_{ij}=c^{\cG}\d_{ij}, \qquad \g_{-\a\b}=c^{\cG}\d_{\a\b}, }
where $c^{\cG}$ is the eigenvalue of the quadratic Casimir in the adjoint
representation. We now write $\a(H_i)=\a_i$,
so that $\a_i$ are orthonormal co-ordinates for the root vectors,
and we perform contractions using the ordinary Dirac delta
function, i.e. $\a.\b=\a_i\b_i$. We also have
\eqn\Aaeb{
[E_{\a},E_{-\a}]=\a.H=\a_iH_i. }
Defining $\psi$ to be the highest root in
$\cG$, we have
\eqn\Aaf{
h^{\cG}={c^{\cG}\over{\psi^2}} }
and also the Freudenthal-De Vries strange formula\ref\FDV{
H. Freudenthal and H. De Vries, {\it Linear Lie Groups\/} (Academic Press,
New York, 1969).},
\eqn\Aag{
24\r^2=h^{\cG}\psi^2\dim\cG }
where $\r$ is defined in Eq. \eIIIga.
We should explain at this point the relation between our present conventions
and notation, and those used in Ref. \jp. In Ref. \jp\ we found it useful
not to restrict ourselves to the case of an orthonormal basis for $\cH$.
We then defined the inner product between roots as follows: first we
defined $H_{\a}$ by requiring $(H_{\a},H)=\a(H)$ for all $H\in\cH$. We then
defined $<\a,\b>=(H_{\a},H_{\b})$. With an orthonormal basis, these
quantities are related to those used here by $\a.H=c^{\cG}H_{\a}$ and
$\a.\b=c^{\cG}<\a,\b>$. As a consequence of Eq. \Aaf, we then have
$<\psi,\psi>={1\over{h^{\cG}}}$.

We now prove Eq. \eIVo. Given a root $\a$, the Weyl reflection
$w(\a)$ corresponding to $\a$ acts on roots $\b$ by
\eqn\Aa{
\b\rightarrow\b^{w(\a)}=\b-2{\b.\a\over{\a.\a}}\a. }
If $\b$ is a root, then $\b^{w(\a)}$ is a root also. Moreover, if $\a$
is a simple root and $\b$ is a positive root (other than $\a$ itself),
then $\b^{w(\a)}$ is also a positive root; for $w(\a)$ only changes the
coefficient of $\a$ in an expansion of $\b$ in terms of simple roots,
and the coefficients of the simple roots in the expansion of $\b^{w(\a)}$
must all be of the same sign. In other words if $\a$ is simple, then $w(\a)$
permutes the positive roots except $\a$ amongst themselves. We thus have, for
a simple root $\a\in\D$,
\eqn\Ab{
\r\rightarrow\r^{w(\a)}=\r-\a, }
where $\r$ is defined in Eq. \eIIIga. Comparing Eqs. \Aa, \Ab, we find
\eqn\Ac{
2{\r.\a\over{\a.\a}}=1,\qquad \a\in\D, }
from which it follows that
\eqn\Ad{
\r=\sum_{\a\in\D}\a^{\pri},  }
with $\a^{\pri}$ as defined in Eq. \eIii.

Now consider $\trho$ as defined in Eq. \eIVna. Weyl reflections in a simple
root of $\ta\in\tD$ permute the positive roots (except $\ta$)
in $\Ph^+$ amongst themselves
and also permute the positive roots (except $\ta$)
in $\tP$ amongst themselves (since $\cG_0$
is a subalgebra). Hence they permute the roots in $\Ph^+\bs\tP$ amongst
themselves (since $\r^{\pri}$ does not contain $\ta$). So we have
\eqn\Ae{
\r^{\pri w(\ta)}=\r^{\pri},\qquad \ta\in\tD  .}
Comparing with Eq. \Aa, we see that
\eqn\Af{
\r^{\pri}.\ta=0,\qquad \ta\in\tD, }
which is what we needed to prove.

We now turn to the proof of Eq. \eIIIk. The Casimir $\cC^{\cG}=\tr[(\cJ^L)^2]$
may be written
\eqn\Ag{
\cC^{\cG}=\sum_{\a\in\Ph^+}(\cJ_{\a}^L\cJ_{-\a}^L+\cJ_{-\a}^L\cJ_{\a}^L)
+\cJ_i^L\cJ_i^L}
where $\cJ_{\a}^L$ is the operator which induces multiplication of $g$ on the
left by $E_{\a}$ and $\cJ_i^L$ is the operator inducing multiplication on
the left by $H_i$. It is easy to see from the definitions \eIIIf\ that
\eqn\Ah{
\cJ_{-\a}^L=X_{-\a}^L;}
however there is no such simple expression for $\cJ_{\a}^L$. It is now
convenient, using
\eqn\Aha{
[\cJ^L_{\a},\cJ^L_{-\a}]=\a_i\cJ^L_i}
 to rewrite Eq. \Ag\ in the form
\eqn\Ai{
\cC^{\cG}=\sum_{\a\in\Ph^+}(2\cJ_{\a}^L\cJ_{-\a}^L-\a_i\cJ_i^L)
+\cJ_i^L\cJ_i^L.}
The reason for writing $\cC^{\cG}$ in this form is that we can now replace the
summation over $\a$ in the first term on the RHS of Eq. \Ai\ by a summation
over $\ta\in\tP$ when $\cC^{\cG}$ acts on $V(\ph_-^{\ta},\ph_+^{\ta},r^i)$.
This is because, from Eq. \Ah\ and the fact, mentioned in
Section 3, that $X_{-\ba}^L$ contains only derivatives with respect to
$\ph_-^{\bar \b}$ for $\bar \b\ge\ba$, it follows that every term in
$\cJ_{\ba}^L\cJ_{-\ba}^L$ contains at least one derivative with respect to
$\ph_-^{\bar \b}$, for some $\bar \b\in\Ph^+\bs\tP$. (This is not the case
for $\cJ_{-\ba}^L\cJ_{\ba}^L$, which can in fact yield terms containing only
a single derivative with respect to $r^i$ for some $i$. This is clear from
Eqs. \Aha\ and \eIIIia.) We now have
\eqn\Ak{
\cC^{\cG}V(\ph_-^{\ta},\ph_+^{\ta},r^i)=(\cC^{\cG_0}-2\r^{\pri}_i\cJ_i^L)
V(\ph_-^{\ta},\ph_+^{\ta},r^i)}
where $\cC^{\cG_0}$ is the Casimir operator for $\cG_0$, and $\r^{\pri}$ is as
defined in Eq. \eIVna. It can readily be shown, using the definitions of the
operators $\cJ$ as generators of multiplication by Lie algebra elements, that
\eqn\Al{
(\cC^{\cG_0}+2\r^{\pri}_i\cJ_i^L)V(\ph_-^{\ta},\ph_+^{\ta},r^i)
=\tr[({\cal O}M_+)g_0M_-g_0^{-1}]}
where
\eqn\Am{
{\cal O}M_+=C^{\cG_0}M_+-\r^{\pri}_i[H_i,M_+],   }
with $C^{\cG_0}$ the ordinary adjoint Casimir for $\cG_0$, i.e.
\eqn\An{
C^{\cG_0}M_+=\sum_{\ta\in\tP}([E_{\ta},[E_{-\ta},M_+]]+
[E_{-\ta},[E_{\ta},M_+]])+[H_i,[H_i,M_+]]. }
It is clear that any $M_+$ with the property \eIxii\ must be a sum of
$E_{\g}$ with $<\d,\g>=0$. From Eq. \eIiiia, any such $\g$ must be of the form
\eqn\Ao{
\g=\ta+\bg_L,}
 where $\ta$ is some root in $\tP$, and $\bg_L$ is a
simple root in $\D\bs\tD$, subject only to the restriction that $\g$ is
indeed a root. Hence, from Eq. \Am, it will be sufficient to prove
\eqn\Ap{
C^{\cG_0}E_{\g}=\r^{\pri}.\g>E_{\g} }
for a root of the form Eq. \Ao. We can write
\eqn\Aq{
E_{\g}=N[E_{\ta},E_{\bg_L}] }
for some constant $N$. The action of the adjoint Casimir $C^{\cG_0}$ commutes
with the action of any elements of $\cG_0$, and hence in particular with
$E_{\ta}$. Hence we have
\eqn\Ar{
C^{\cG_0}E_{\g}=N[E_{\ta},C^{\cG_0}E_{\bg_L}]. }
We can write
\eqn\As{
C^{\cG_0}E_{\bg_L}=2\sum_{\ta\in\tP}([E_{\ta},[E_{-\ta},E_{\bg_L}]]+
[H_i,[H_i,E_{\bg_L}]]
-2\trho_i[H_i,E_{\bg_L}],  }
where $\trho$ is as defined in Eq. \eIVna. Since $\bg_L$ is a simple root, the
first term on the RHS vanishes, and we have
\eqn\At{
C^{\cG_0}E_{\bg_L}=(\bg_L-2\trho).\bg_LE_{\bg_L}.  }
Using Eqs. \Ad\ and \eIii, we can write
\eqn\Au{
\bg_L^2=2\r.\bg_L, }
and hence we have, from Eqs. \At, \Au\ and \eIVna,
\eqn\Av{
C^{\cG_0}E_{\bg_L}=\r^{\pri}.\bg_LE_{\bg_L}.  }
Finally, combining Eqs. \Af, \Ao, \Aq, \Ar\ and \Av, we obtain Eq. \Ap.
\listrefs
\bye